\documentclass[aps,prb,twocolumn,showpacs,superscriptaddress,floatfix]{revtex4-1}
\usepackage{amsmath,amssymb}
\usepackage{graphicx}
\usepackage{epsfig}
\usepackage{color}
\usepackage{rotating}
\usepackage{bm}
\usepackage{appendix}
\usepackage{setspace}
\begin{document}
\title{Anisotropic superconductivity and elongated vortices with unusual bound states in quasi-one-dimensional nickel-bismuth compounds}
\author{Wen-Lin Wang}
\author{Yi-Min Zhang}
\author{Yan-Feng Lv}
\author{Hao Ding}
\affiliation{State Key Laboratory of Low-Dimensional Quantum Physics, Department of Physics, Tsinghua University, Beijing 100084, China}
\author{Lili Wang}
\author{Wei Li}
\author{Ke He}
\author{Can-Li Song}
\email[]{clsong07@mail.tsinghua.edu.cn}
\author{Xu-Cun Ma}
\email[]{xucunma@mail.tsinghua.edu.cn}
\author{Qi-Kun Xue}
\email[]{qkxue@mail.tsinghua.edu.cn}
\affiliation{State Key Laboratory of Low-Dimensional Quantum Physics, Department of Physics, Tsinghua University, Beijing 100084, China}
\affiliation{Collaborative Innovation Center of Quantum Matter, Beijing 100084, China}
\date{\today}

\begin{abstract}
We report low temperature scanning tunneling microscopy and spectroscopy studies of Ni-Bi films grown by molecular beam epitaxy. Highly anisotropic and twofold symmetric superconducting gaps are revealed in two distinct composites, Bi-rich NiBi$_3$ and near-equimolar Ni$_x$Bi, both sharing quasi-one-dimensional crystal structure. We further reveal axially elongated vortices in both phases, but Caroli-de Gennes-Matricon states solely within the vortex cores of NiBi$_3$. Intriguingly, although the localized bound state splits energetically off at a finite distance $\sim$ 10 nm away from a vortex center along the minor axis of elliptic vortex, no splitting is found along the major axis. We attribute the elongated vortices and unusual vortex behaviors to the combined effects of twofold superconducting gap and Fermi velocity. The findings provide a comprehensive understanding of the electron pairing and vortex matter in quasi-one-dimensional superconductors.
\end{abstract}
%\pacs{74.55.+v, 74.25.Ha, 68.37.Ef, 74.70.Ad}
\maketitle
\begin{spacing}{1}
\section{\label{sec:Intro}Introduction}
Superconductivity in reduced dimensionality has long been of interest in condensed matter physics, particularly given that most superconductors with relatively higher transition temperature $T_\textrm{c}$, regardless of whether they are unconventional cuprates \cite{bednorz1986possible} and iron pnictides \cite{kamihara2008iron} or conventional MgB$_2$ \cite{nagamatsu2001superconductivity}, have quasi-two-dimensional (2D) layered structure. It becomes more interesting insofar after the interface superconductivity with an unexpectedly high $T_\textrm{c}$ $ >$ 65 K in single-layer FeSe films grown on SrTiO$_3$ was recently discovered \cite{qing2012interface}. In the last three decades, much has been learned about the novel pairing phenomena and exotic phenomena in quasi-2D superconductors. As the dimensionality is further reduced to quasi-one-dimensional (1D) case, however, how the electrons would pair up and behave is little explored. This has been hindered primarily by the rare occurrence of superconductivity in quasi-1D electron systems. Thus far, only a few quasi-1D compounds, such as the Bechgaard salts (TMTSF)$_2$X (X = PF$_6$, ClO$_4$) \cite{jerome1980superconductivity, Bechgaard1981zero}, Li$_{0.9}$Mo$_6$O$_{17}$ \cite{Mercure2012upper}, Ta$_4$Pd$_3$Te$_{16}$ \cite{jiao2014superconductivity} and the recently-discovered chromium pnictides \textit{A}$_2$Cr$_3$As$_3$ (\textit{A} = K, Rb and Cs) \cite{superconductivity2015Bao}, are found to be superconducting. A comprehensive understanding of electron pairing and vortex states in quasi-1D systems is currently lacking.

Recently, owing to its availability of cleavage plane, the quasi-1D Ta$_4$Pd$_3$Te$_{16}$ superconductor with an anisotropy of about 2.5 \cite {Nodal2015Pan} was investigated on the atomic scale by means of scanning tunneling microscopy (STM) \cite{scanning2015fan, du2015anisotropic}, but leading to several obvious controversies. First, the measured superconducting energy gaps were distinctively accounted for by isotropic \cite{scanning2015fan} or anisotropic \cite{du2015anisotropic} electron pair functions. Second, although the elongated magnetic vortices were consistently observed, there exists great ambiguity as regards the vortex core states. Neither the observed double-peak feature around the vortex centers \cite{scanning2015fan} nor the non-split zero-bias conductance (ZBC) peaks when away from the vortex center \cite{du2015anisotropic} appears contradictory with previous reports, in which the ZBC peaks caused by vortex bound states\cite{caroli1964bound} often split off energetically starting from the vortex center in both anisotropic \cite{Hess1990vortex, Guillamon2008intrinsic} and isotropic superconductors \cite{Guillam2008superconducting}. A concern over the electron pairing and vortex states has therefore emerged in quasi-1D superconductors. In this study, we reveal the anisotropic superconductivity and unusual vortex states in two quasi-1D Ni-Bi films grown by molecular beam epitaxy (MBE), which exhibits a comparable anisotropy with Ta$_4$Pd$_3$Te$_{16}$ \cite{fujimori2000superconducting}, and provide a detailed microscopic view of quasi-1D superconductors.

\section{\label{sec:Exper}Experiment}

Our experiments were carried out in an ultrahigh vacuum cryogenic STM system (Unisoku), which is connected with a MBE chamber for \textit{in-situ} sample preparation. The base pressure of both chambers is better than 1.0 $\times$ $10^{-10}$ Torr. Nb-doped ($\sim$ 0.05 wt\%) SrTiO$_3$(001) substrates were cleaned by annealing at $1250^{\circ}$C for 20 min. High-quality Ni-Bi epitaxial films were grown on SrTiO$_3$(001) at $\sim$ $280^{\circ}$C$-300^{\circ}$C by thermal evaporation of high-purity Ni (99.994\%) and Bi (99.997\%) sources from two standard Knudsen cells. Details of the sample preparation are given in Appendix A. The temperatures of Ni and Bi sources were respectively set at $1200^{\circ}$C and $495^{\circ}$C, resulting in a growth rate of approximately 0.1 nanometer per minute. The optimal growth was achieved under Bi-rich condition, bearing a similarity to that for $\beta$-Bi$_2$Pd films \cite{lv2017experimental}. Polycrystalline PtIr tips were cleaned by electron beam heating and calibrated prior to data collection at 0.4 K, unless otherwise specified. Tunneling conductance spectra and ZBC maps were measured by means of standard lock-in technique with a small bias modulation of 0.1 mV at 913 Hz.

\end{spacing}

\section{\label{sec:Res}Results}
\subsection{\label{sec:sample}Surface structure}
Based on their mole ratio, the alloys of Ni and Bi can be crystalized into three distinct phases \cite{nash1985bi}, with the crystal structures and lattice parameters summarized in Table I. The typical STM topographies of the three Ni-Bi phases, which can coexist on the as-grown films, are shown in Figs.\ 1(a)-1(c). Illustrated in Fig.\ 1(a) is the first stoichiometric NiAs-type NiBi phase with the space group P63/\textit{mmc} [Fig.\ 1(d)], which presents a hexagonal supercell. The adjacent bright spheres are spaced 7.2 $\pm$ 0.2 ${\textrm{\AA}}$ apart, consistent with a reconstructed NiBi(0001)-$(\sqrt{3}\times\sqrt{3})R30^\textrm{o}$ surface.

\begingroup
\begin{table}[h]
\center
\caption{\label{tab:crystal}Lattice parameters and structure of Ni-Bi phases.}
\begin{footnotesize}
\scalebox{1.00}{
\begin{tabular}{ l l c c c }
  \hline\hline
 &  & NiBi & Ni$_x$Bi & NiBi$_3$   \\
  \hline
   &\textit{a}(${\textrm{\AA}}$) & 4.08 & 14.12 & 8.88   \\
   &\textit{b}(${\textrm{\AA}}$) & 4.08 & 8.15  & 4.11   \\
   &\textit{c}(${\textrm{\AA}}$) & 5.36 & 5.32 & 11.48   \\
   &Space group & P63/\textit{mmc} & F12/\textit{m}1 & P\textit{nma}   \\
   &Structure& hexagonal  & pseudo-orthorhombic & orthorhombic  \\
  \hline
\end{tabular}}
\end{footnotesize}
\end{table}
\endgroup

As the Ni/Bi ratio slightly departs from unity, however, more specialized studies have previously found pseudo-orthorhombic Ni-Bi superstructure with a space group of F12/\textit{m}1 \cite{ruck1999kristallstruktur, lidin2000incommensurately}, in which displaced and occupational modulations of Bi and Ni atoms lead to quasi-1D structure along the $b$ axis. This near-equimolar Ni$_x$Bi phase, as represented in Fig.\ 1(b), actually exhibits the expected in-plane lattice parameters: \textit{a} = 14.12 ${\textrm{\AA}}$, \textit{b} = 8.15 ${\textrm{\AA}}$ [Fig.\ 1(e)]. Note that lattice discontinuities are always found to run along the $b$ axis [Fig.\ 1(b)] and mainly originate from the off-stoichiometry created Ni/Bi vacancies. These vacancies appear beneficial that one can easily tell the atoms along the 1D chain apart and measure the lattice parameter of \textit{b} = 8.15 ${\textrm{\AA}}$.

In Fig.\ 1(c), we show the third phase that appears predominantly at a lower growth temperature of $280^{\circ}$C. The in-plane lattice constants are measured as \textit{a} = 8.17 ${\textrm{\AA}}$ and \textit{b} = 4.13 ${\textrm{\AA}}$ from the inserted high resolution STM image in Fig.\ 1(c), which matches with the stoichiometric NiBi$_3$ phase \cite{fujimori2000superconducting, kumar2011physical, pineiro2011possible, Herr2011structure, Surface2012zhu, Silva2013Super}. The experimental lattice constants for all the three Ni-Bi phases are further confirmed by Fast Fourier transform (FFT) images of the three Ni-Bi species [Figs.\ 1(g)-1(i)], with the white arrows indicating the 2D reciprocal vectors $a^*$ and $b^*$. By measuring the lengths of $a^*$ and $b^*$ vectors, we can calculate the lattice parameters in real space based on their reciprocal relationship. For example, in Ni$_x$Bi, we measured the lengths of $a^*$ to be 0.71/nm, giving rise to a lattice constant $a$ = 14.08 $\textrm{\AA}$, close to the theoretical value of 14.14 \AA. We can also obtain the lattice constants for the other two phases, which are in line with the values obtained directly from the real space and match well with the theoretical expectation. Notably, the Bragg spots along the vector $b^*$ is too weak to be resolved in Ni$_x$Bi and one can not tell the lattice constant $b$ from the FFT image. Fortunately, we measure the lattice parameter of $a$ = 8.15 $\textrm{\AA}$ from the real space STM image, as discussed above.

Although some early studies indicated the coexistence of superconductivity and ferromagnetism in NiBi$_3$ \cite{pineiro2011possible, Herr2011structure, Surface2012zhu}, the recent work reveals that it might be amorphous Ni residuals to bear the responsibility for the magnetic signal \cite{Silva2013Super}. As drawn in Fig.\ 1(f), the NiBi$_3$ could be best described as a packing of NiBi$_3$ rods. Within a single rod, the bonding is dominated by Ni-Ni and Ni-Bi interactions, while the Bi-Bi bonding appears significantly weak. Such a structural confinement provides another quasi-1D system \cite{fujimori2000superconducting, Surface2012zhu}, similar to the near-equimolar Ni$_x$Bi.

\begin{figure}[t]
\includegraphics[width=1\columnwidth]{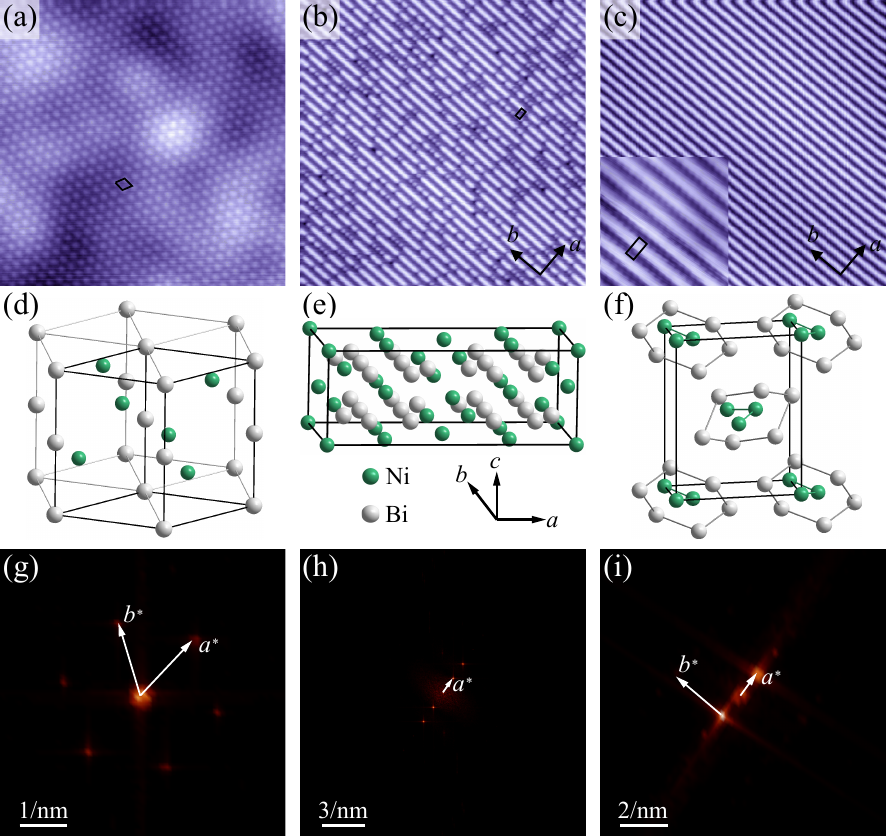}
\caption{(color online) (a)-(c) STM topographies of the as-grown Ni-Bi films presenting three distinct phases, i.e.\ (a) hexagonal NiBi (\textit{V} = 50 mV, \textit{I} = 300 pA, 20 nm $\times$ 20 nm), (b) pseudo-orthorhombic Ni$_x$Bi (\textit{V} = 50 mV, \textit{I} = 30 pA, 45 nm $\times$ 45 nm) and (c) NiBi$_3$ (\textit{V} = 100 mV, \textit{I} = 30 pA, 26.7 nm $\times$ 26.7 nm). The white spots in (a) might originate from subsurface defects. The inset of (c) shows a higher resolution STM image of NiBi$_3$ (\textit{V} = 100 mV, \textit{I} = 30 pA, 6 nm $\times$ 6 nm). The black rhombus and rectangles mark the unit cells of three Ni-Bi phases. (d)-(f) Schematic crystal structures of NiBi, Ni$_x$Bi and NiBi$_3$, with their unit cells marked by the thick black lines. The \textit{a}, \textit{b} and \textit{c} axes are oriented along the crystal directions of the two orthorhombic phases in (b) and (c). Note that the atom modulations in near-equimolar Ni$_x$Bi are not displayed in (e). (g-i) FFT images of (a) hexagonal NiBi, (b) near-equimolar Ni$_x$Bi and (c) NiBi$_3$ films, respectively.
}
\end{figure}

\subsection{\label{sec:super}Anistropic superconductivity}

\begin{figure*}[t]
\includegraphics[width=2\columnwidth]{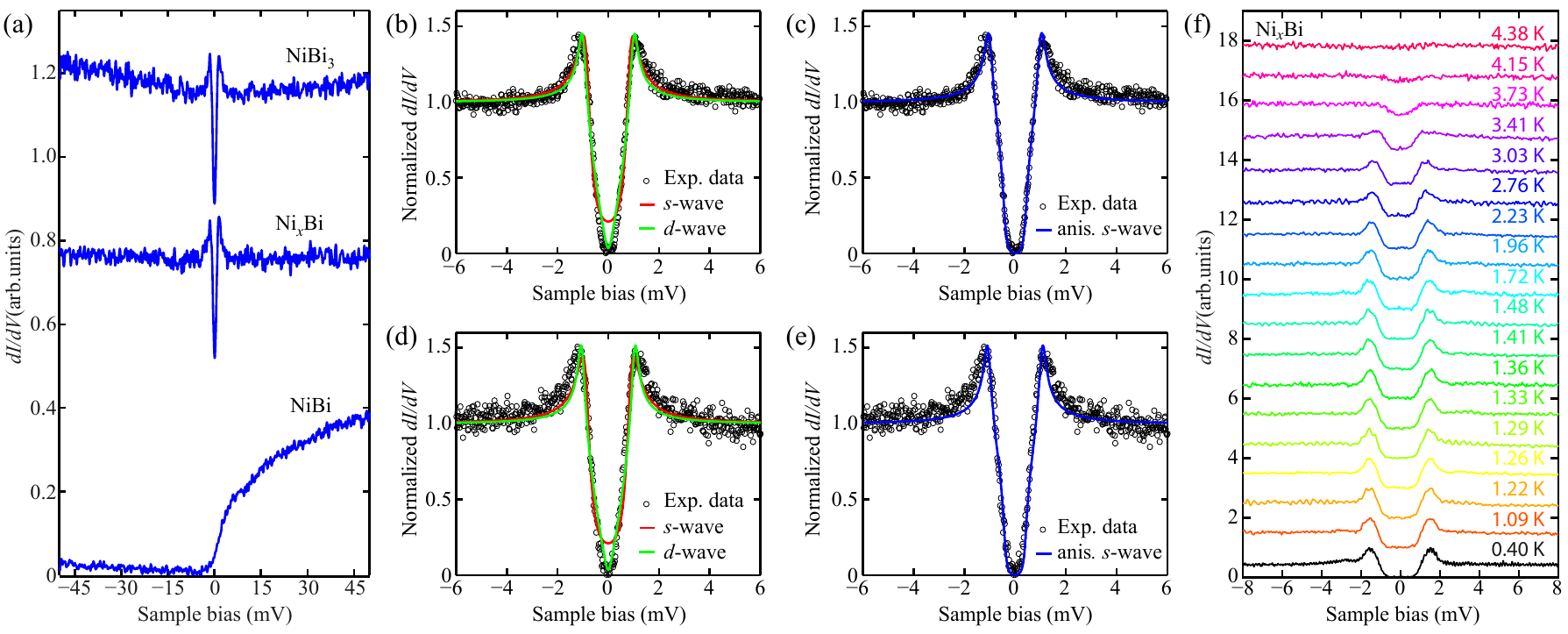}
\caption{(color online) (a) Differential conductance tunneling \textit{dI/dV} spectra measured on Ni-Bi films. The curves on Ni$_x$Bi and NiBi$_3$ have been vertically shifted for clarity. Set point: \textit{V} = 50 mV, \textit{I} = 100 pA. (b)-(e) Normalized \textit{dI/dV} spectrum (black circles) on (b, c) NiBi$_3$ and (d, e) Ni$_x$Bi. The coloured lines show the best fits of experimental data to the Dynes model with \textit{s}-wave, \textit{d}-wave and anisotropic \textit{s}-wave gap functions. The normalization was performed by dividing the raw \textit{dI/dV} spectra by their backgrounds, extracted from a linear fit to the conductance beyond the superconducting gaps. Set point: \textit{V} = 10 mV, (b, c) \textit{I} = 200 pA, (d, e) \textit{I} = 100 pA. (f) Temperature-dependent \textit{dI/dV} spectra of Ni$_x$Bi films with a superconducting tip, revealing the disappearance of superconductivity at 4.38 K. The tunneling gap is stabilized at $V$ = 10 mV and $I$ = 100 pA.
}
\end{figure*}

\begin{figure*}[t]
\includegraphics[width=2\columnwidth]{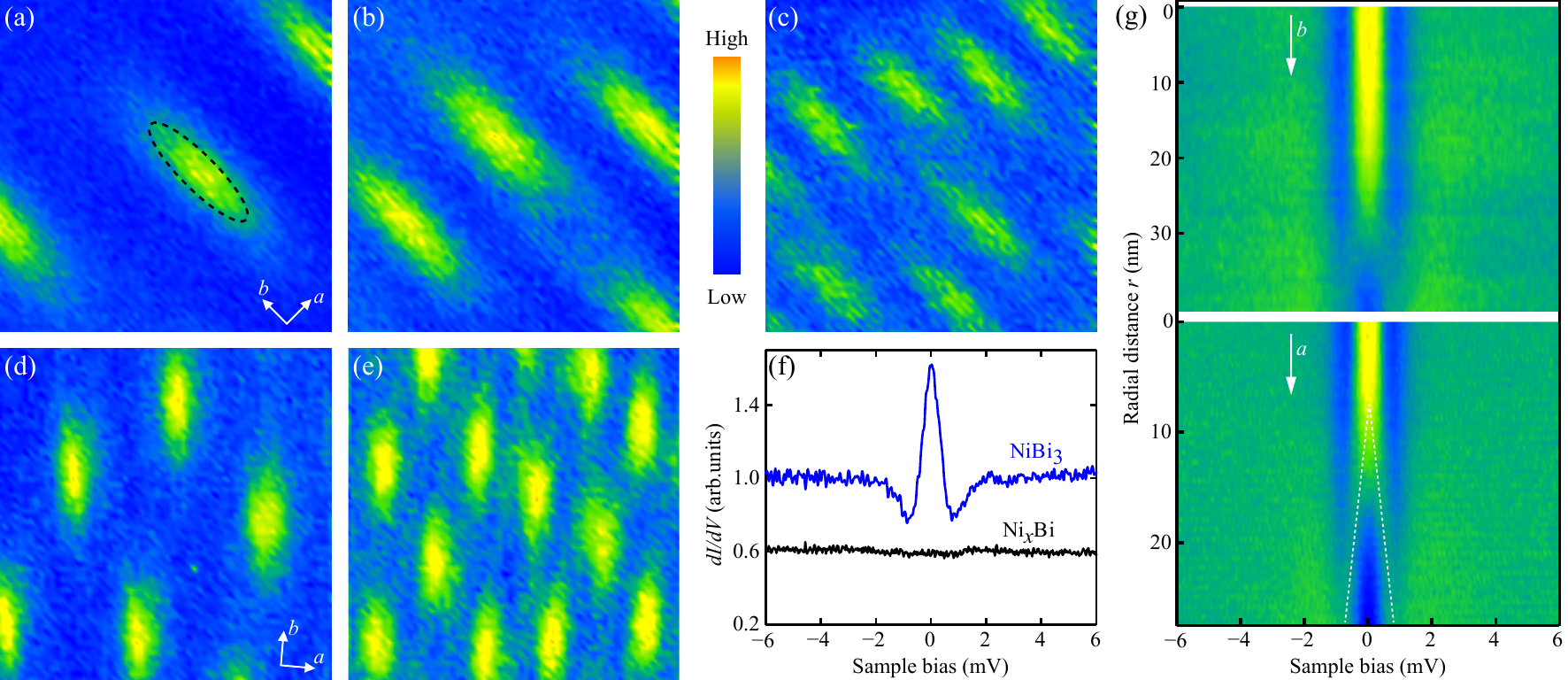}
\caption{(color online) (a)-(e) Zero-bias-conductance maps on (a-c) NiBi$_3$ (270 nm $\times$ 270 nm) and (d, e) Ni$_x$Bi (350 nm $\times$ 350 nm) films under various magnetic fields, revealing strongly axially elongated vortices. The black ellipse in (a) illustrates how unidirectional a single vortex is. The fields have magnitudes of 0.03 T in (a), 0.1 T in (b, d) and 0.2 T in (c, e), respectively. (f) STS curves on the center of vortex cores. (g) Dependences of \textit{dI/dV} spectra of NiBi$_3$ on the radial distance $r$ from a vortex center and crystallographic direction. The dotted lines mark the splitting of ZBC peaks along the \textit{a} crystal axis. The tunneling gap is set at \textit{V} = 10 mV, \textit{I} = 100 pA except for (f, g) \textit{I} = 200 pA.
}
\end{figure*}

Scanning tunneling spectroscopy (STS) probes quasiparticle density of states (DOS) and provides key information of superconductivity. To eliminate possible disturbance from SrTiO$_3$ substrate, we here focus on thick Ni-Bi films with a thickness greater than 10 nm. Plotted in Fig.\ 2(a) are the large-energy-scale STS data at 0.4 K. Despite a sudden DOS change around the Fermi level ($E_F$), the NiBi has no signature of superconductivity. In contrast, both of NiBi$_3$ and Ni$_x$Bi films exhibit symmetric gaps with respect to $E_F$. The gaps are suppressed and eventually vanish at elevated temperatures, indicating their origin from superconductivity. This thus provides two platforms to examine directly the electron pairing of quasi-1D superconductors by STS.

In order to gain insight into the pairing gap, we have collected and normalized the smaller-energy-scale \textit{dI/dV} spectra, shown in Figs.\ 2(b)-2(e) as black empty symbols. We use tentatively the Dynes model with an effective energy broadening $\Gamma$ \cite{Direct1978dynes}, namely $dI/dV \cong \textrm{Re}[(E-i\Gamma)/\sqrt{(E-i\Gamma)^2-\triangle^2}]$, to fit the experimental data with different categories of superconducting gaps, including isotropic \textit{s}-wave gap, \textit{d}-wave gap and anisotropic \textit{s}-wave gap. The results for optimized fits are color coded in Figs.\ 2(b)-2(e). It is evident that the isotropic \textit{s}-wave scenario, marked by the red lines in Figs.\ 2(b) and 2(d), fails completely to track the low-energy DOS around $E_F$. Although the green solid lines with \textit{d}-wave gaps can more reasonably follow the STS curves as a whole, it gives rise to somewhat too sharp V-shaped feature near the bottom, as compared to the STS data. We thus consider the scenario of anisotropic \textit{s}-wave gap function to simulate the data. As justified below, the twofold symmetric gap $\Delta(\theta)=\Delta_1+\Delta_2\textrm{cos}(2\theta)$ rather than four-fold symmetric gap is favorable, although both functions lead to the same gap structure. As illustrated in Figs.\ 2(c) and 2(e), the anisotropic \textit{s}-wave gaps interpret the experimental STS data quite nicely. This results in the gap function $\Delta_{\textrm{NiBi}_\textrm{3}} (\textrm{meV}) =0.72+0.30 \textrm{cos}(2\theta)$ and $\Delta_{\textrm{Ni}_x\textrm{Bi}} (\textrm{meV}) =0.75+0.32\textrm{cos}(2\theta)$ for NiBi$_3$ and Ni$_x$Bi, respectively. The gap maximum $\Delta_{\textrm{max}}$ for NiBi$_3$ (Ni$_x$Bi) is estimated to be 1.02 meV (1.07 meV), while the gap minimum $\Delta_{\textrm{min}}$ is 0.42 meV for both films. This suggests that despite large anisotropy there exists no node in the gap function.

The nodeless pairing is further confirmed by picking up a nanometer-sized Ni-Bi superconducting film on the end of STM tip and performing the superconductor-insulator-superconductor tunneling spectrum, which allows for higher energy resolution \cite{ji2008high}. Figure 2(f) plots such STS spectra as a function of temperature, which were acquired on the Ni$_x$Bi films. At the lower temperatures, the spectra reveal vanishing DOS over a finite energy range near $E_F$. This gives the convincing evidence of no nodes involved in the superconducting gap function of Ni-Bi films investigated. At elevated temperatures, the superconducting gaps are progressively suppressed and completely vanishes above 4.38 K. This suggests that the $T_\textrm{c}$ of Ni$_x$Bi studied lies between 4.15 K and 4.38 K ($\sim$ 4.2 K), consistent with the reported value of 4.25 K \cite{fujimori2000superconducting}. We also measured the \textit{dI/dV} spectrum on NiBi$_3$ and revealed the disappearance of superconducting gap at an elevated temperature of 4.3 K, consistent with the expected $T_\textrm{c}$ = 4.06 K \cite{fujimori2000superconducting}. Therefore, we estimate $T_\textrm{c}$ $\sim$ 4.2 K for Ni$_x$Bi epitaxial films, although more careful experiments are needed to determine the $T_\textrm{c}$ for the NiBi$_3$ films. Using the gap maxima $\Delta_{\textrm{max}}$, we extract the reduced gap $2\Delta_{\textrm{max}}/k_BT_c$ is $\sim$ 5.8 for Ni$_x$Bi, suggesting strong coupling superconductivity in Ni-Bi compounds. One might argue that multi-band effects could be involved and a mixture of isotropic \textit{s}-wave and $d$-wave gaps can also fit the STS data. However, a recent study from magneto-resistance measurements has revealed the dominance of only one band below 60 K in NiBi$_3$ \cite{Surface2012zhu}. The Ni-Bi films are thus most likely characteristic of anisotropic single superconducting gap.

\subsection{\label{sec:mag}Magnetic vortices and unusual core states}

\begin{figure}[b]
\includegraphics[width=0.7\columnwidth]{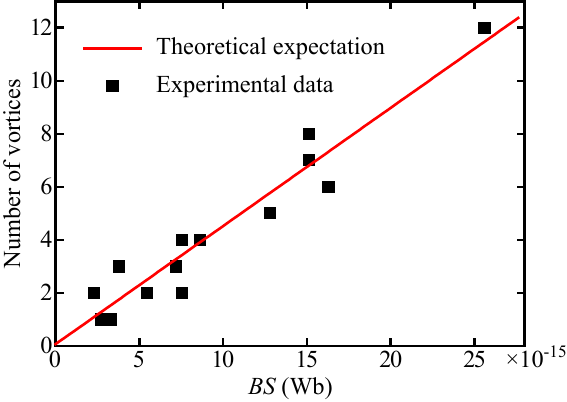}
\caption{Number of quantized magnetic vortices as a function of $BS$, with $B$ and $S$ representing the applied field and the area of ZBC maps holding the vortices, respectively. The red line corresponds to a liner fit of the experimental data.
}
\end{figure}

Magnetic vortex imaging gains further insight into gap structure and vortex states of superconductors. Figures 3(a)-(e) show the spatial ZBC maps under various magnetic fields, applied perpendicularly to the NiBi$_3$ [Figs.\ 3(a)-(c)] and Ni$_x$Bi [Figs.\ 3(d) and 3(e)] sample surfaces. The yellow stripe-like regions have large ZBC due to the suppressed superconductivity and correspond to individual vortex cores, which are remarkably elongated along the $b$ axis. The density of vortices is found to increase linearly with the applied magnetic field [Fig.\ 4], as anticipated.  The average flux per vortex is calculated to be $(2.09\pm0.16)\times10^{-15}$ Wb, consistent with a single magnetic flux quantum, $2.07\times10^{-15}$ Wb. Here the elongated vortex cores can be intuitively accounted for by the directional dependence of coherence lengths $\xi$. To calculate $\xi$, in Fig.\ 5(a) we extract the radial and directional dependence of ZBC in NiBi$_3$ and fit them by the Ginzburg-Landau expression for superconducting order parameter: ZBC($r$)=ZBC$_\propto$+(1$-$ZBC$_\propto$)(1$-$exp($-$$r/\sqrt{2}\xi$)) with ZBC$_\propto$ representing the ZBC far away from the vortex core \cite{vortex2002Eskildsen}. Figure 5(b) plots the angular dependence of $\xi$, from which we estimate the ratio $\xi_b/\xi_a$ of as high as 3.9 in NiBi$_3$. This value appears larger than the anisotropy of Fermi velocity $v_{Fb}/v_{Fa}$ = 2.6 \cite{fujimori2000superconducting}. Since $\xi$ is cooperatively given by $v_{F}$ and $\Delta$ via $\xi = \hbar v_F/\pi \Delta$, this supports the above-claimed anisotropy in $\Delta$. Meanwhile, the pairing gap must be of twofold symmetry with gap minima along the \textit{b} axis. Otherwise, the vortex core should be of fourfold or modulated fourfold symmetry. This thus rules out the possible involvement of \textit{d}-wave gap in quasi-1D NiBi$_3$. Such a rule might happen in Ni$_x$Bi as well (blue symbols in Fig.\ 5(b)). Here The ratio $\xi_b/\xi_a$ is estimated to be about 3.8 and larger than the gap anisotropy of 2.55, suggesting that the Fermi velocity $v_{F}$ is different along the $a$ and $b$ axes. This behaves analogous to that observed in NiBi$_3$. Moreover, the occurrence of gap minimum along the \textit{b} axis is consistent with the small DOS there, because the DOS scales inversely with $v_{F}$ \cite{Nakai2002reentrant}, which in our case appears larger along the \textit{b} axis than that along the \textit{a} axis. Notably the above results are subject to confirmation by the angular-dependent measurements of magnetoresistance and upper critical field $H_{\textrm{c2}}$. However, unfortunately, these parameters are not currently available, since the Ni-Bi samples studied here are only a few hundred nanometers in size and not large enough for a macroscopic investigation

\begin{figure}[t]
\includegraphics[width=1\columnwidth]{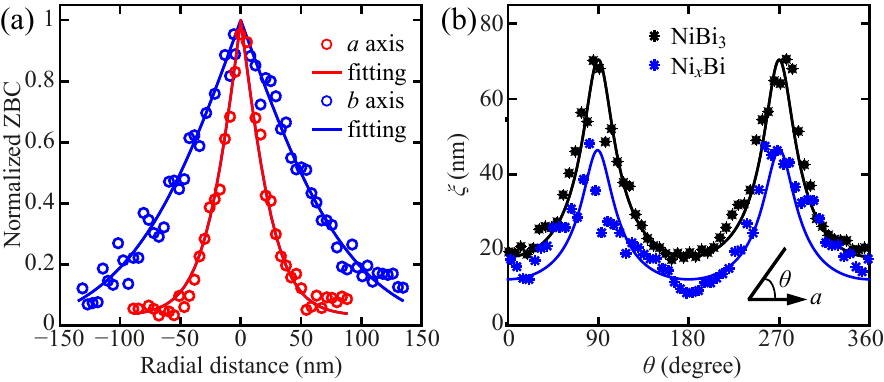}
\caption{(color online) (a, b) Radial and directional dependence of normalized ZBC around a single magnetic vortex along the \textit{a} (red spots) and \textit{b} (blue spots) axes on NiBi$_3$ films, respectively. Solid lines show the theoretical fits of ZBC in NiBi$_3$. (b) Directional dependence of coherence length $\xi$ in NiBi$_3$ (black symbols) and Ni$_x$Bi (blue symbols) films. The angle $\theta$ is measured against the \textit{a} axis. The solid lines are guides to eyes.
}
\end{figure}

The most remarkable finding stands out by exploring the low-energy quasiparticle excitations in the vicinity of vortices. In Fig.\ 3(f), we show the STS data at the centers of vortices in NiBi$_3$ and Ni$_x$Bi. Although the NiBi$_3$ holds a pronounced ZBC peak, no ZBC peak exists in Ni$_x$Bi. Such a distinction might originate from the fact that off-stoichiometry-increased scattering significantly reduces the electron mean free path $\ell$ and pushes the Ni$_x$Bi superconductor into the dirty region ($\ell<\xi$), where the constructive interference of repeated Andreev scatterings responsible for the ZBC peak can no longer sustain within the vortex cores. Conversely, the presence of ZBC peak means a longer $\ell$ than $\xi$ (clean limit) in NiBi$_3$, which coincides with its high stoichiometry nature. Figure 3(g) shows the dependence of tunneling \textit{dI/dV} spectra on the radial distance $r$ away from a vortex center in NiBi$_3$. Quite strikingly, the ZBC peaks evolve quite differently between along the \textit{a} and \textit{b} axes. Although the ZBC peak firstly decays and then begins to split into two symmetric branches in energy at a distance of $r \sim$ 10 nm along the \textit{a} axis (bottom panel), it exhibits no observable splitting along the \textit{b} axis (upper panel). We emphasize that this unusual finding is inherent to superconducting NiBi$_3$, irrespective of the field of view we studied. This is well confirmed in Fig.\ 6. In a different field of view, we observe the identical behaviors of the vortex core states.

 \begin{figure}[t]
\includegraphics[width=1\columnwidth]{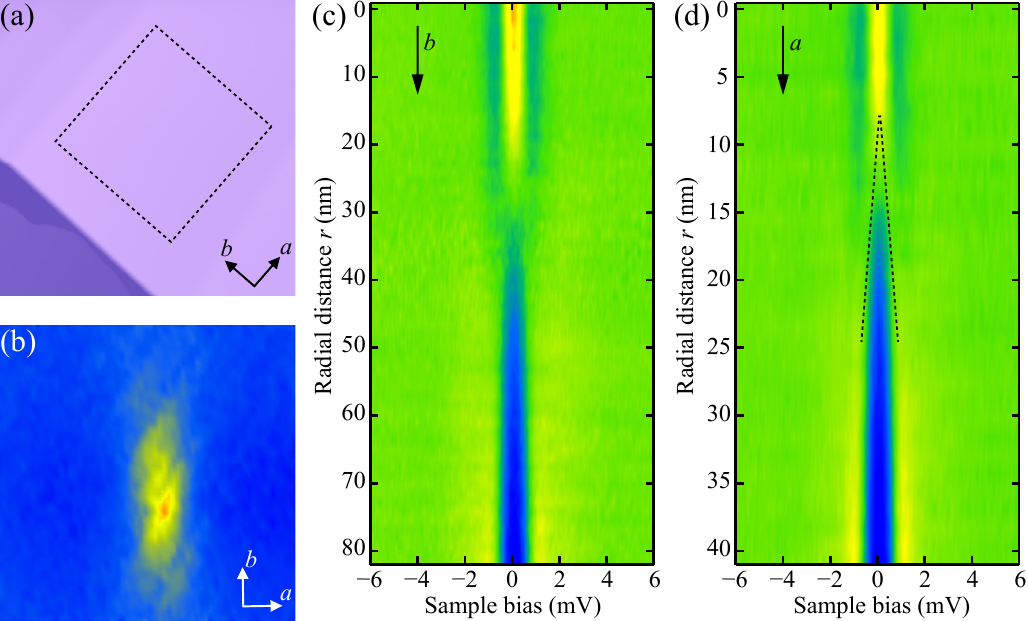}
\caption{(a) STM topographic image taken on another field of view of NiBi$_3$ films ($V$ = 3.5 V, $I$ = 30 pA, 300 nm $\times$ 300 nm). (b) Zero-bias-conductance map ($V$ = 10 mV, $I$ = 100 pA, 168 nm $\times$ 168 nm) taken on the square dashes-marked region in (a) at 0.1 T. (c, d) Spatial and directional dependence of \textit{dI/dV} spectra ($V$ = 10 mV, $I$ = 200 pA). The radial distance $r$ is measured from the cortex center. The dotted lines in (d) mark the ZBC peak splitting along the \textit{a} crystal axis, while no splitting is observed in (c).
}
\end{figure}

\section{\label{sec:Dis}Discussions}
The unprecedented variations of vortex-induced low-energy excitations differ substantially from the common Caroli-de Gennes-Matricon (CdGM) vortex bound states \cite{caroli1964bound}, which were often observed to split at sites quite close to the vortex center \cite{song2011direct, Hess1990vortex}. The retarded splitting of such peak around vortices was rarely found in topological insulator Bi$_2$Te$_3$/NbSe$_2$ hybrid structure and interpreted as the involvement of Majorana zero modes \cite{Xu2015experimental, Kawakami2015evolution}. The zero-energy Majorana mode comes into play together with the CdGM states and will delay the splitting of ZBC peaks around vortices. This might be possible for NiBi$_3$ because the Ni/Bi bilayer films, which inevitably contains NiBi$_3$ phase during the sample preparation \cite{siva2015spontaneous}, exhibit nontrivial superconducting states potentially harboring MZMs at vortices \cite{xin2015possible, wang2016observation, gong2017time}.

However, a more reasonable explanation might be that the observed vortex-core-state anomaly in Fig.\ 3(g) and Fig.\ 6 come solely from CdGM states in quasi-1D compounds with twofold superconducting properties. In the theoretical framework of CdGM states, the low-energy quasiparticle excitations around vortices consist of a series of discrete bound states, with the spacing of energy levels proportional to $\Delta$ and $1/v_F$ \cite{Hayashi1998low}. Along the \textit{b} axis of quasi-1D NiBi$_3$, the smaller $\Delta$ and larger $v_F$ conspire to lead to a small energy spacing of the bound states, which is difficult to be resolved due to the thermal smearing of 0.4 K in experiment. On the other hand, the energy spacing of bound states along the \textit{a} axis (large $\Delta$ and small $v_F$) might be large enough to be discernible at a finite distance away from the vortex center (the bottom panel of Fig.\ 3(g)). This scenario reasonably explains our findings above and can help understand the direction-dependent vortex core states in Ta$_4$Pd$_3$Te$_{16}$ as well \cite{scanning2015fan, du2015anisotropic}. Further experiment at a lower temperature might help fully understand the mechanism behind the exotic vortex states observed in Fig.\ 3(g).

Finally we comment on vortex arrangement in quasi-1D superconductors. For isotropic $s$-wave superconductors, and when pinning is not quite effective, vortices are packed into a hexagonal lattice. On the other hand, square vortex lattice with the sides along the directions of $\Delta$ or $v_F$ minima, can be formed in superconductors with fourfold symmetric properties \cite{Sakata2000imaging, Nakai2002reentrant}. In quasi-1D compounds with twofold superconductiving state, however, the vortices are more distorted and could be roughly described by an oblique lattice [Figs.\ 3(c)-(e)], although the NiBi$_3$ films are clean and have negligible vortex pinning. Such an oblique vortex lattice seems generic for superconductors with quasi-1D features, such as YBa$_{2}$Cu$_{3}$O$_{7-\delta}$ with \textit{ab}-plane crystal anisotropy \cite{Maggio1995direct}, FeSe with strong electronic nematicity \cite{song2011direct} and Ta$_{4}$Pd$_{3}$Te$_{16}$ with chain-like structure \cite{scanning2015fan, du2015anisotropic}.  This might be due to the trade-off between the direction-dependent vortex-vortex interactions for elliptic vortices and the closet packing of vortices.

\section{\label{sec:Sum}Summary}
In summary, our detailed STM/STS study of Ni-Bi epitaxial films has developed a comprehensive microscopic picture of quasi-1D superconductors. First, we have explicitly revealed by conducting tunneling spectrum that quasi-1D superconductors are characteristic of twofold symmetric gap functions. Second, we have visualized elongated vortices, and more significantly revealed their (as well as the unusual CdGM vortex bound states) origin from the combined effects of twofold superconducting pairing and Fermi velocity in quasi-1D superconductors. Third, we have shown by spatially resolved spectroscopy that the CdGM bound states behave sharply different along the major and minor axes of elliptic vortices in quasi-1D superconductors. This allows for a reasonable explanation for the discrepancy of CdGM states in Ta$_4$Pd$_3$Te$_{16}$, and provides a complete microscopic view of vortex matter in quasi-1D superconductors.

\begin{acknowledgments}
This work was financially supported by the Ministry of Science and Technology of China (2017YFA0304600, 2016YFA0301004) and National Science Foundation of China. C. L. S. acknowledges supports from the National Thousand-Young-Talents Program and the Tsinghua University Initiative Scientific Research Program.
\end{acknowledgments}

\begin{appendix}
\section{Sample preparation}

 We prepared Ni-Bi epitaxial films on SrTiO$_3$(001) substrate by using Bi-rich condition. A deficiency of Bi element during the growth was found to result in polycrystalline Ni clusters. To grow crystalline Ni-Bi films, we have set the temperatures of Ni and Bi sources at $1200^{\circ}$C and $490^{\circ}$$\sim$ $ 500^{\circ}$C, respectively, which leads to a Bi/Ni flux ratio of around 1.5$\sim$2.2. A larger flux ratio of Bi/Ni leads to many unwanted small rods (most probably the excess Bi), in particular at a lower substrate temperature $T_{\textrm{sub}}$ [Fig.\ 7(a)].

 \begin{figure}[b]
\includegraphics[width=1\columnwidth]{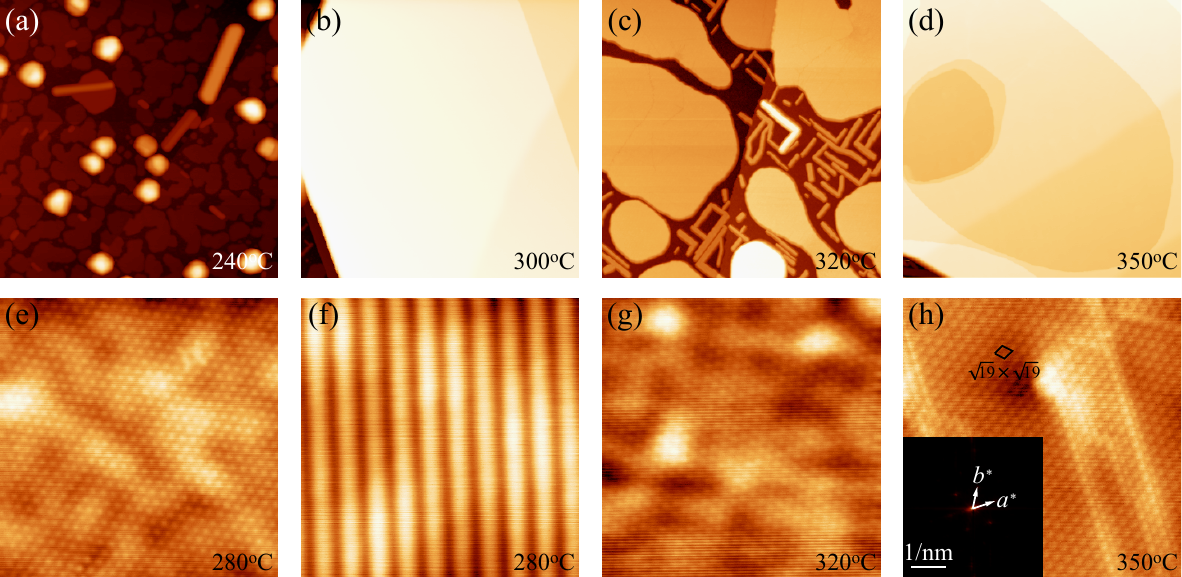}
\caption{ (a-d) Morphology ($V$ = 3.5 V, $I$ = 25 pA, 500 nm $\times$ 500 nm except for d 430 nm $\times$ 430 nm) and (e-h) zoom-in STM images of as-grown Ni-Bi films at various $T_{\textrm{sub}}$ as indicated. Tunneling conditions and image size: (e) $V$ = 50 mV, $I$ = 30 pA, 18 nm $\times$ 18 nm; (f) $V$ = 50 mV, $I$ = 30 pA, 13 nm $\times$ 13 nm; (g) $V$ = 10 mV, $I$ = 30 pA, 18 nm $\times$ 18 nm; and (h) $V$ = 100 mV, $I$ = 26 pA, 45 nm $\times$ 45 nm.
}
\end{figure}

Figure 7 illustrates the STM morphologies and the corresponding atom resolution STM images of as-grown Ni-Bi films. The Ni-Bi compounds grow typically in a three-dimensional island mode. At the elevated substrate temperature $T_{\textrm{sub}}>280^{\circ}$C [Figs.\ 7(b-d)], the epitaxial Ni-Bi films with a lateral size of hundreds of nanometers could be found, which well satisfy our requirements for STM studies. A closer scrutiny of atomically-resolved STM images reveal that although the stoichiometric NiBi, near-equimolar Ni$_x$Bi and Bi-rich NiBi$_3$ phases coexist in an intermediate substrate temperature $T_{\textrm{sub}}$ = $280^{\circ}$$\sim$ $300^{\circ}$C, a higher or lower $T_{\textrm{sub}}$ leads to the exclusive hexagonal phase with various reconstructed surface structrues [Figs.\ 7(e),\ 7(g) and 7(h)]. For example, at $T_{\textrm{sub}}$ = $350^{\circ}$C, the frequently observed surface exhibits a hexagonal superstructure with an extremely large period [Fig.\ 7(h)]. Based the inserted Fast fourier transform (FFT) image in Fig.\ 7(h), the periodicity of the superstructure is estimated to be approximately 1.79 nm, matching excellently with a reconstructed NiBi(0001)-$(\sqrt{19}\times\sqrt{19})$ surface. The absences of Bi-rich NiBi$_3$ and Ni$_x$Bi phases at the higher $T_{\textrm{sub}}$ are understandable, since the sticking coefficient of Bi on SrTiO$_3$ is significantly small at the higher $T_{\textrm{sub}}$ \cite{lv2017experimental}, leading to a Bi-deficient environment in which only the Bi-poor hexagonal phase is formed. On the other hand, we hypothesize that the reason why no Bi-rich NiBi$_3$ and Ni$_x$Bi phase exist at the lower $T_{\textrm{sub}}$ might be their relatively higher formation energy  than NiBi. In order to obtain the NiBi$_3$ and Ni$_x$Bi phases, a sufficiently high $T_{\textrm{sub}}$ ($>280^{\circ}$) is thus required.

\end{appendix}

%\bibliography{NiBi}

%

\end{document}